\documentclass[%
reprint,amsmath,amssymb,twocolumn,
 aps,
 prb,
 ]{revtex4}
\usepackage{hyperref}
\usepackage{graphicx}% Include figure files
\usepackage{dcolumn}% Align table columns on decimal point
\usepackage{bm}

\begin{document}

\title{Bulk and Surface Nucleation Processes in Ag$_2$S Conductance Switches}

\author{M. Morales-Mas\'{\i}s}
\email{morales@physics.leidenuniv.nl}
\affiliation{%
 Kamerlingh Onnes Laboratorium, Universiteit Leiden, PO Box 9504, 2300 RA Leiden, The Netherlands\\
}%

\author{S.J. van der Molen}%
\affiliation{%
 Kamerlingh Onnes Laboratorium, Universiteit Leiden, PO Box 9504, 2300 RA Leiden, The Netherlands\\
}%

\author{T. Hasegawa}
\affiliation{WPI Center for Materials Nanoarchitectonics, National Institute for Materials Science, 1-1 Namiki, Tsukuba, Ibaraki 305-0044, Japan\\
}%

\author{J. M. van Ruitenbeek}
\affiliation{%
 Kamerlingh Onnes Laboratorium, Universiteit Leiden, PO Box 9504, 2300 RA Leiden, The Netherlands\\
}%

\begin{abstract}

We studied metallic Ag formation inside and on the surface of Ag$_2$S thin films, induced by the electric field created with a STM tip. Two clear regimes were observed: cluster formation on the surface at low bias voltages, and full conductance switching at higher bias voltages (V $>$ 70mV). The bias voltage at which this transition is observed is in agreement with the known threshold voltage for conductance switching at room temperature. We propose a model for the cluster formation at low bias voltage. Scaling of the measured data with the proposed model indicates that the process takes place near steady state, but depends on the STM tip geometry. The growth of the clusters is confirmed by tip retraction measurements and topography scans. This study provides improved understanding of the physical mechanisms that drive conductance switching in solid electrolyte memristive devices.

\end{abstract}

\maketitle

\section{Introduction}

The interest to study metal cluster formation in chalcogenide materials, comes from the prospect of building  nanometer-scale conductance switches with the configuration Metal/Chalcogenide/Metal (M/C/M). In this configuration, changes in the conductance are induced by the application of a voltage between the two metal electrodes. Depending on the chalcogenide material, the conductance switching is caused by phase transitions \cite{Wuttig2007, Lencer2008} or ionic transport processes \cite{Waser2007, Waser2009, Hino2011, Kozicki2010}.\\

The amplitude of the voltage necessary to switch an M/C/M system from the low conductance state to the high conductance state, has a threshold value ($V_{th}$). This value is the voltage at which the energy barrier for nucleation of a metallic phase inside the chalcogenide material is overcome. The threshold voltages are strongly dependent on the chalcogenide material, electrode material, and geometry of the metal-chalcogenide-metal cell \cite{Geresdi2011, Nayak2010, Karpov2008, Waser2009, tamura2}.\\

In the case of ionic transport, conductance switching is due to the formation and dissolution of metallic filaments between the electrodes. The formation and dissolution of such nanoscale metallic filaments is driven by the polarity and amplitude of the voltage applied \cite{Waser2007, Morales-Masis2009}. One of the distinguishing properties of ionic switches is the fact that the diffusion constant of ions at room temperature can be very large, which is an advantage in view of low energy operation (low voltages to switch on and off), but a disadvantage in view of a short memory retention time.\\

We focus on the study of conductance switching due to ion transport and we use Ag$_2$S as the model material. Our interest lies in the understanding of the physical processes that drive nucleation and filament formation, including threshold voltages and growth kinetics. Previously, we studied the transport processes within Ag$_2$S thin films, which occur under steady state conditions, i.e. before switching \cite{Morales-Masis2010}. We used a conducting AFM tip brought in contact with the Ag$_2$S sample. For Ag nucleation inside the Ag$_2$S film giving full conductance switching, a critical supersaturation of Ag ions inside the Ag$_2$S film is needed. This phenomenon only occurs at bias voltages above 70mV \footnote{This value depends on the non-stoichiometry of Ag$_{2+\delta}$S. The value of 70 mV applies for a sample for which the Ag cation concentration is fixed by an intimate contact to a Ag metal reservoir.}, which is defined as the threshold voltage for switching at room temperature. In this article we confirm experimentally this threshold voltage for nucleation inside the film. We furthermore show that when starting from a tunneling gap between the top electrode and the sample, Ag clusters can be grown between the tip and the Ag$_2$S surface at voltages significantly lower than the observed threshold voltage for switching. At voltages above the threshold, a rapid nucleation inside the Ag$_2$S thin film is observed and the sample then exhibits full conductance switching. The two processes can be separated by properly preparing the initial conditions. 

\section{Experimental procedure}

The samples consist of Ag(200nm)/Ag$_2$S(200nm) layers deposited on a Si substrate. The Ag film is sputtered on a Si(100) substrate covered with a native oxide layer. On top of the Ag layer, the Ag$_2$S film is grown by sputtering of Ag in a Ar/H$_2$S plasma. The sample preparation is described in more detail in Ref. \onlinecite{Morales-Masis2010}. The surface roughness of the sample is approximately 30nm.\\

As the top electrode, we use an STM tip which is manually cut from either a Pt or PtIr wire. The measurements are performed at room temperature in a JEOL UHV STM/AFM system with a base pressure of 1 x 10$^{-9}$ mbar. In this STM setup, the tunneling voltage is applied to the sample while the tip is connected to ground.\\

For the measurements we have connected an external data acquisition card (DAQ) from National Instruments to the JEOL STM controller. Our measurement software allows the simultaneous control of two outputs (bias voltage, piezo voltage) and the monitoring and recording of two inputs (current and Z piezo voltage). Hence, the measurements are fully controlled through Labview independent of the STM controller.\\

\section{Conditions for vacuum tunneling on $\text{Ag}_2\text{S}$}

When performing STM measurements on Ag$_2$S, two important points must be considered: 1. Ag$_{2}$S is an n-type semiconductor, with a band gap ranging between 0.6 and 1.2 eV \cite{Kashida2003, Wang2008}. 2. Ag$_2$S has both ionic and electronic mobile charges. The ionic mobile charges (Ag$^+$-ions) act as n-type dopants within the Ag$_2$S film. Therefore, a local accumulation of ions in the sample causes changes in the local conductivity and band gap of the film. In our experiments, we apply a voltage between the STM tip (top contact) and the Ag layer (bottom contact), creating a strongly localized electric field near the tip. If a positive sample bias is applied, the mobile Ag$^+$-ions in the Ag$_2$S will move towards the region closest to the tip, increasing the local conductivity and lowering the band gap. Due to the points mentioned above, the apparent sample resistance can vary depending on the values of the tunneling gaps and applied voltages.\\

\begin{figure}[ht]
  \begin{center}
        \includegraphics[width=9cm]{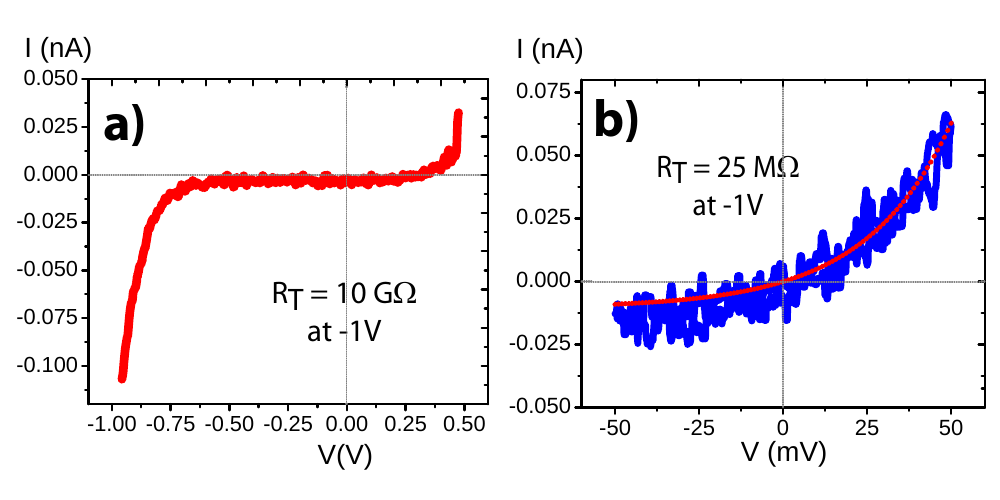}      % data from 181010-controller computer
  \caption{Current-Voltage characteristics of the Ag$_2$S thin film at two different tunneling gaps. a. At 10 G$\Omega$ (tunneling gap defined at -1V), the IV curve indicates the presence of the band gap in the Ag$_2$S sample. b. At 25 M$\Omega$ (tunneling gap defined at -1V), the IV curve shows an exponential increase in the current at positive bias, due to the accumulation of Ag close to the tip for close proximity of the tip to the sample. The red curve is a fit of the IV curve from which we calculate the effective top contact radius as 0.8nm .} 
   \label{1to100mV}
  \end{center}
\end{figure}

We measured IV curves to characterize our sample and confirm the characteristics of the band gap and the effect of the mobile ions. A typical IV curve, measured at the tunneling gap of 10 G$\Omega$, is shown in Figure \ref{1to100mV}a. The IV curve is asymmetric, with a flat region between approximately -600mV and +400mV. This flat region is the result of the band gap of the sample. At the negative bias the rapid increase in conductance below -600mV is associated with the valence band edge. At the positive bias side above +400mV the rise of the curve is related to the position of the conduction band edge. The position of the conduction band edge can be influenced by the n-type doping due to the accumulation of Ag$^+$-ions.\\

The effects of n-type doping become more important when the tunneling gap is reduced. When working at tunneling gaps with a resistance in the range of several M$\Omega$, a large part of the voltage drops across the sample (the sample resistance for a Pt atomic point contact with the Ag$_2$S sample is $>$ 3 G$\Omega$ for voltages below the band edge), and therefore the activation of ions to move towards the tip is higher than for large tunneling gaps. Figure \ref{1to100mV}b presents an IV curve measured at a tunneling gap of 25 M$\Omega$, set at a bias of -1V. The IV curve shows an exponential increase in the current at the positive bias side, at a voltage much lower than the band gap edge. This IV curve resembles our previous measurements \cite{Morales-Masis2010} and confirms the prediction by Hebb \cite{HEBB1952} and Wagner \cite{WAGNER1953}, of an accumulation of Ag$^+$-ions resulting in an enhanced electronic conductivity and an asymmetric IV curve with an exponential increase in the current at positive sample bias. Additionally, fitting the exponential IV curve with the Hebb-Wagner formalism (red curve in Figure \ref{1to100mV}b), allow us to calculate the top contact radius. We obtained an effective contact radius of 0.8 nm, indicating the close proximity, or touching, of the tip to the sample. \\

In order to assure that the tip is not in contact with the Ag$_2$S film, we choose to work with tunneling resistance between 10 G$\Omega$ and 1 G$\Omega$, which is set at a bias below the band edge at a voltage of -1V. We use positive sample bias to activate the ion mobility and Ag growth.\\

\section{Nucleation and Conductance Switching}

\begin{figure}[ht]
  \begin{center}
        \includegraphics[width=8.5cm]{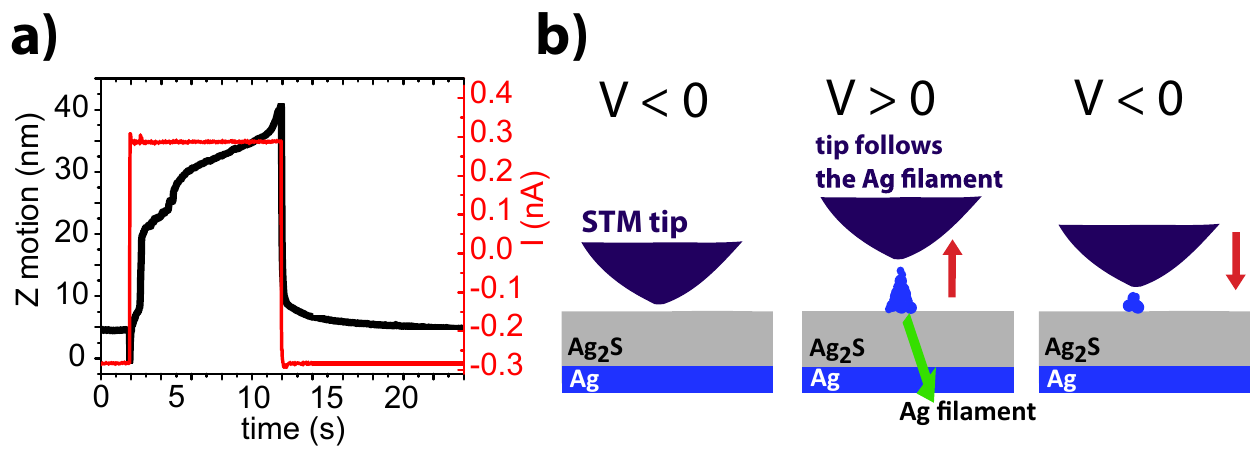}          % data from 071010
      \caption{a. Z piezo displacement (black curve) and tunneling current (red curve) as a function of time. The growth of the Ag cluster was activated with +20mV bias. Keeping the feedback on allow us to follow the growth of the cluster in the Z direction. b. Schematic of the filament growth on the surface of the Ag$_2$S sample.}
    \label{tracefON}
  \end{center}
\end{figure}

Our initial goal was to measure the growth rate of the Ag cluster (or filament), by following its growth with the STM tip, where the tip-sample (or tip-cluster) distance was kept constant by using the proper feedback parameters \cite{Terabe2}. Figure \ref{tracefON} is an example of such a measurement. For this specific trace, a tunneling gap is first defined at -1V sample bias and 0.3 nA tunneling current. Two seconds after starting the measurement, the bias voltage is stepped from -1V to +20mV (change in sign of the current in figure \ref{tracefON}). Almost immediately, an increase in the voltage output of the Z piezo is measured, which continues until the bias voltage is switched back to -1V. The increase of the Z piezo voltage indicates the total displacement of the tip away from the sample. In 10s we measure a total displacement of 36 nm. The tip displacement, therefore indicates the growth of a protrusion on the surface of the sample. We expect this protrusion to be composed of Ag atoms, due to the reduction of Ag$^+$-ions at the surface of the Ag$_2$S film \cite{Xu2010, Terabe2}. \\

Although qualitatively successful, this type of measurements has the disadvantage that tip contact with the sample cannot be avoided when the voltage is stepped from -1V (tunneling voltage) to the positive voltage to induced the filament growth. Because of the large resistance of the sample at low voltages, initially the voltage drops over the sample, not over the tunnel gap. The feedback responds to the resulting drop in the current by pushing the tip into contact with the sample (note the sharp spike to a lower Z position at the start in Figure \ref{tracefON}a). At the same time a filament starts growing. In many occasions the cluster grows much faster than the feedback speed of our STM, causing large oscillations in the Z position and tunneling current. Because of this, the results show large variations in the growth rate, and it is not possible to accurately determine the dependence of the growth rate on the bias voltage.\\

As a different approach we decided to allow contact of the cluster with the tip, by switching off the feedback of the STM and measuring the evolution of the current with time (instead of the Z piezo displacement). At the start of the measurements, the tunneling gap is fixed at -1V sample bias and 1nA tunneling current. With a tunneling gap thus defined, the feedback is switched off and a positive bias voltage is applied to the sample. The positive bias voltage is kept constant while the current is being recorded as a function of time.\\

\begin{figure}[h]
  \begin{center}
          \includegraphics[width=7cm]{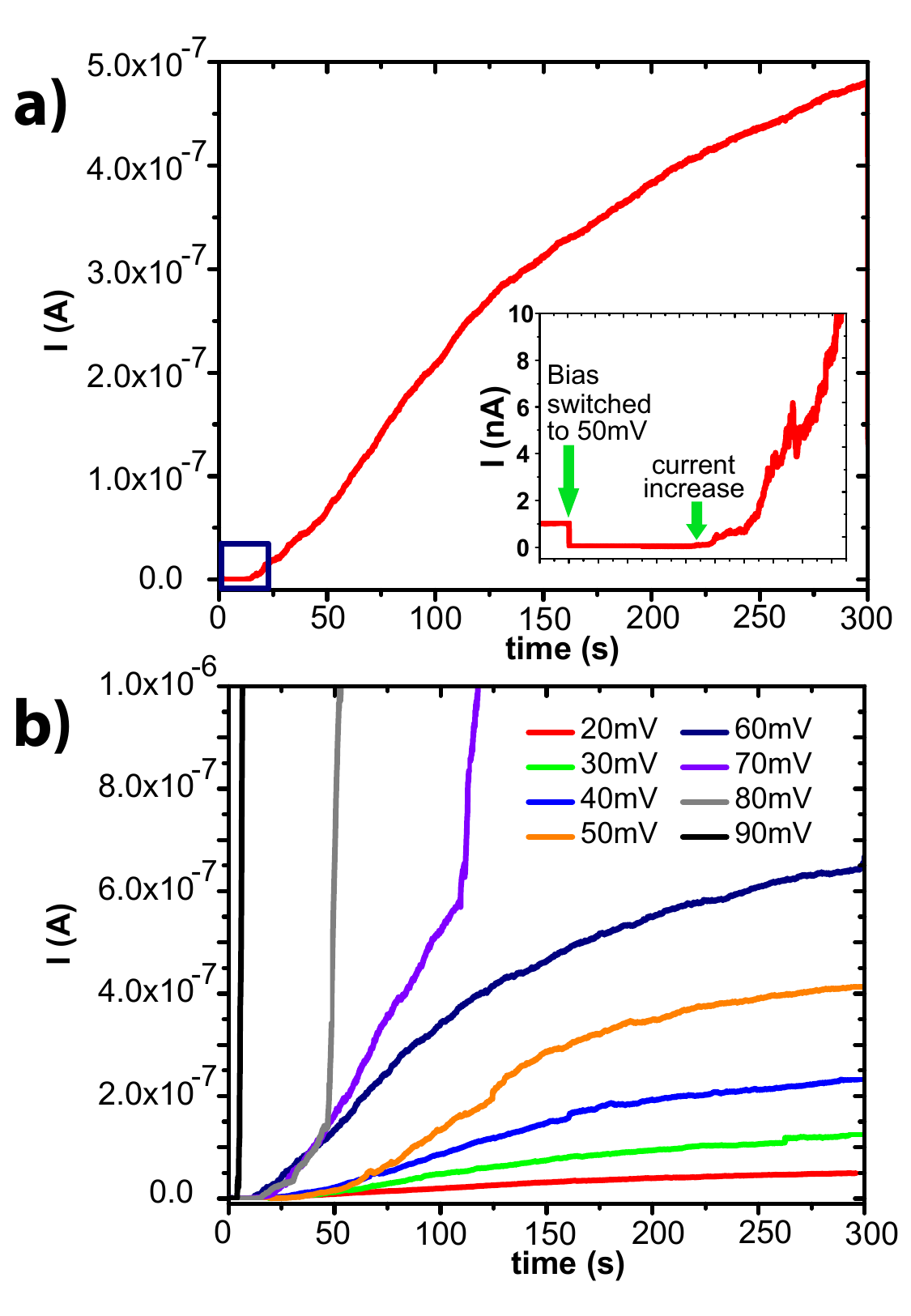}      		 % data from 270311
      \caption{a. Current evolution observed when a step in the voltage from -1V to +50mV is applied to the sample. The inset shows the first seconds of the trace. b. Traces measured at bias voltages from +20mV up to +90mV}
    \label{trace50mV150211}
  \end{center}
\end{figure}

Figure \ref{trace50mV150211}a presents a typical measurement. The measurement is performed as follows: at t=0, the feedback is switched off, and we wait 2 seconds to confirm that the tip-sample distance remains constant. At t=2s the bias voltage $V$ is switched from -1V to +50mV. Immediately we observe a drop in the current, as expected from the high resistance of the sample at low bias voltages. A certain time elapses (induction time) before the current starts increasing (see inset in Figure \ref{trace50mV150211}a). The current increases until the bias voltage is switch back to -1V.  \\
%For larger times the current follows approximately a square root of time relation. \\

The same measurement was performed for bias voltages, $V$, ranging from +20mV to +90mV (Figure \ref{trace50mV150211}b). In this set of data, for the traces taken with +20mV up to +60mV we observe the same behavior as described for Figure \ref{trace50mV150211}a, with a systematic increase in the growth rate as a function of the applied bias voltage.\\

At +70mV a particular curve is observed: during the first 100 seconds it resembles the previous traces. However, this behavior is interrupted by a sudden increase of the current growth rate. The current increase is such that it quickly saturates the current amplifier. A similar behavior is observed at +80mV, with the main difference that the strong increase in current is observed at an earlier time. At +90mV, the increase of the current occurs almost immediately after the positive voltage is applied. \\

\begin{figure}[h]
  \begin{center}
  			  \includegraphics[width=8cm]{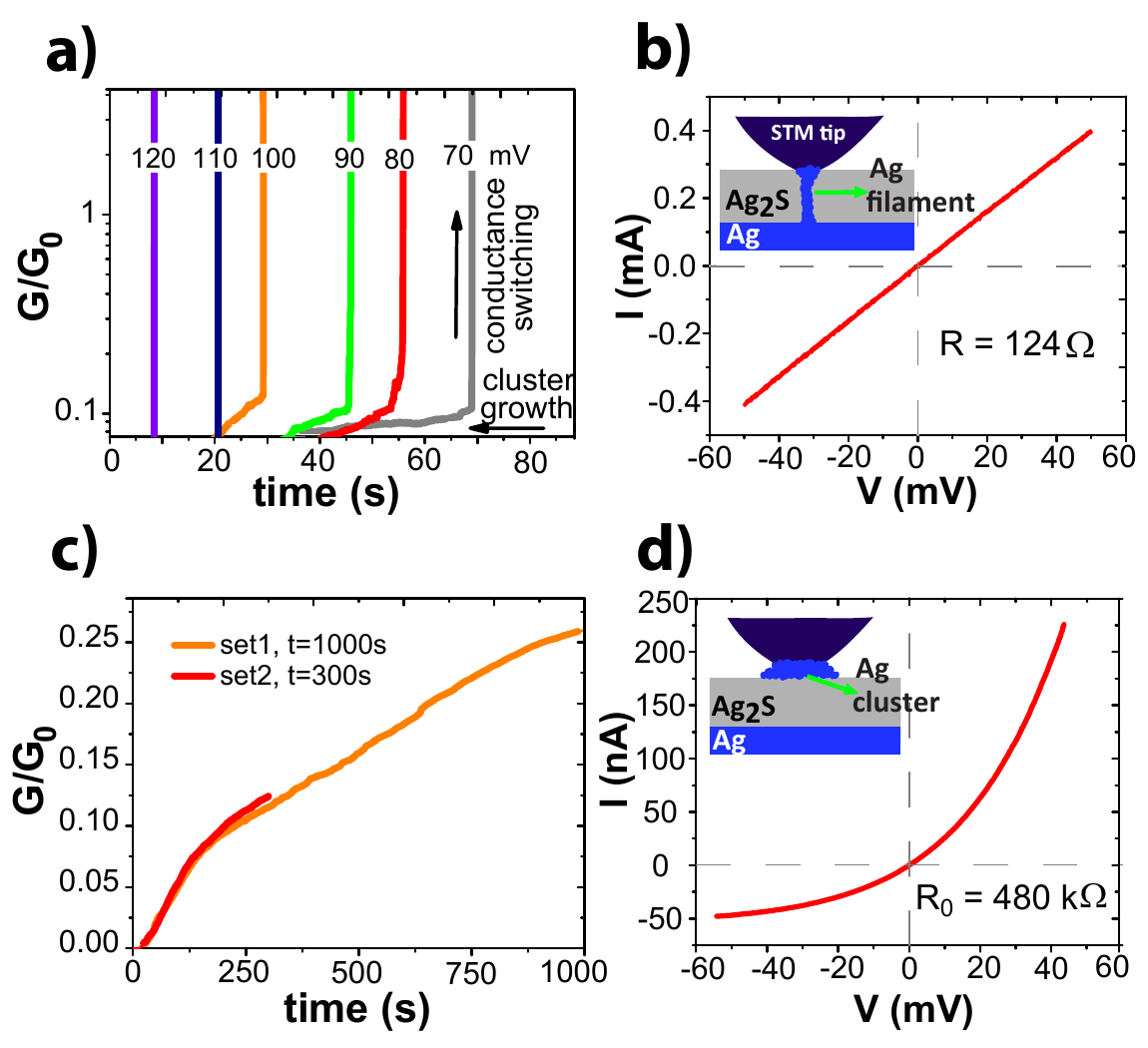}	
	\caption{a. Conductance switching induced at bias voltages $\geq$ +70mV. The plot shows the increase of the conductance above 1$G_0$ and a systematic decrease in the induction time with bias. b. IV curve measured in the high conductance state of the sample, indicating metallic behavior. c. Conductance trace measured for 1000s, applying a bias voltage of +50mV. The trace in red presents another trace measured at +50 mV for 300s. The trace does not show a sudden increase of the conductance for full switching. d. IV curve measured at t = 60s of a trace measured with +50mV, demonstrating semi-conducting behavior dominated by the bias induced changes in ion concentration in the Ag$_2$S film \cite{Morales-Masis2010}.}
		\label{4}
  \end{center}
\end{figure}

The measurements presented in Figure \ref{trace50mV150211}b, suggest the presence of two distinct physical processes. Figure \ref{4} presents further measurements performed in order to distinguish the two processes. Figure \ref{4}a presents a set of traces of conductance (G) as a function of time, obtained by applying bias voltages of +70mV and higher. Each run the sample conductance increases sharply, reaching values larger than 1$G_0$, where $G_0 = 2e^2/h = 77 \mu$S is the conductance quantum. We observe a systematic decrease of the induction time with bias. At $G > 1G_0$, the presence of a metallic Ag filament connecting the bottom electrode with the STM tip is expected. In the traces presented in Figure \ref{4}a, the final conductances (not shown in the plot) reach values as high as hundreds $G_0$. In order to test for metallic conductance, we switch the sample to the high conductance state with +110 mV and measure an IV curve (Figure \ref{4}b). The resulting curve is linear, with a slope indicating a resistance of 124 $\Omega$, in this example. This confirms the metallic behavior of the sample which we attribute to the formation of a Ag filament inside the Ag$_2$S film.\\

To verify that the filament formation does not occur in the low voltage regime (from 0 to +60mV) we measured traces for times much longer than the 300s typically used. Figure \ref{4}c shows one of the traces measured by applying a bias voltage of +50mV for 1000s. The trace is plotted as the conductance (G/G$_0$) vs time. It is clear from the figure, that there is no sudden increase in the current and that the sample, even after 1000s, has a conductance lower than 1$G_0$, although it is not much smaller. At this point, we performed IV measurements to probe the current-voltage behavior at this stage. Figure \ref{4}b presents the resulting non-linear IV curve, which confirms that the conductance is dominated by the dynamic doping behavior of the Ag$_2$S semiconductor.\\ 

Two processes are clearly observed from the behavior of the traces of current vs time: Ag surface nucleation, and Ag filament formation inside the Ag$_2$S film. At room temperature we demonstrate that this transition occurs at 70mV. It is important to mention that this threshold voltage varies with temperature and with the electrode material used. We have used Ag as the bottom electrode, which fixes the chemical potential at the Ag/Ag$_2$S interface \cite{SCHMALZRIED1980} and therefore all changes occur near the Pt tip. In previous studies, we observed that if we use Pt as the bottom as well as for the top contact the threshold voltage to full conductance switching increases up to 200mV \cite{Morales-Masis2009}.\\

In the current versus time traces the increase of the current is linked to the growth of a Ag cluster on the sample surface, giving rise to an increasing contact size to the Ag$_2$S film. Because the feedback of the STM is turned off when the positive bias voltage is applied, the tip does not change its position and the growth of a cluster will occur between the STM tip and the sample surface. As the gap between tip and surface is only of the order 1nm, the cluster will make contact with the STM tip at the very start of the process. As long as the positive bias is applied the cluster will continue growing. The growth is expected to be in the radial direction increasing the contact area with the Ag$_2$S surface and filling the volume between the tip and the surface. These two points are tested further in the following.\\

\begin{figure}[h]
  \begin{center}
        \includegraphics[width=9cm]{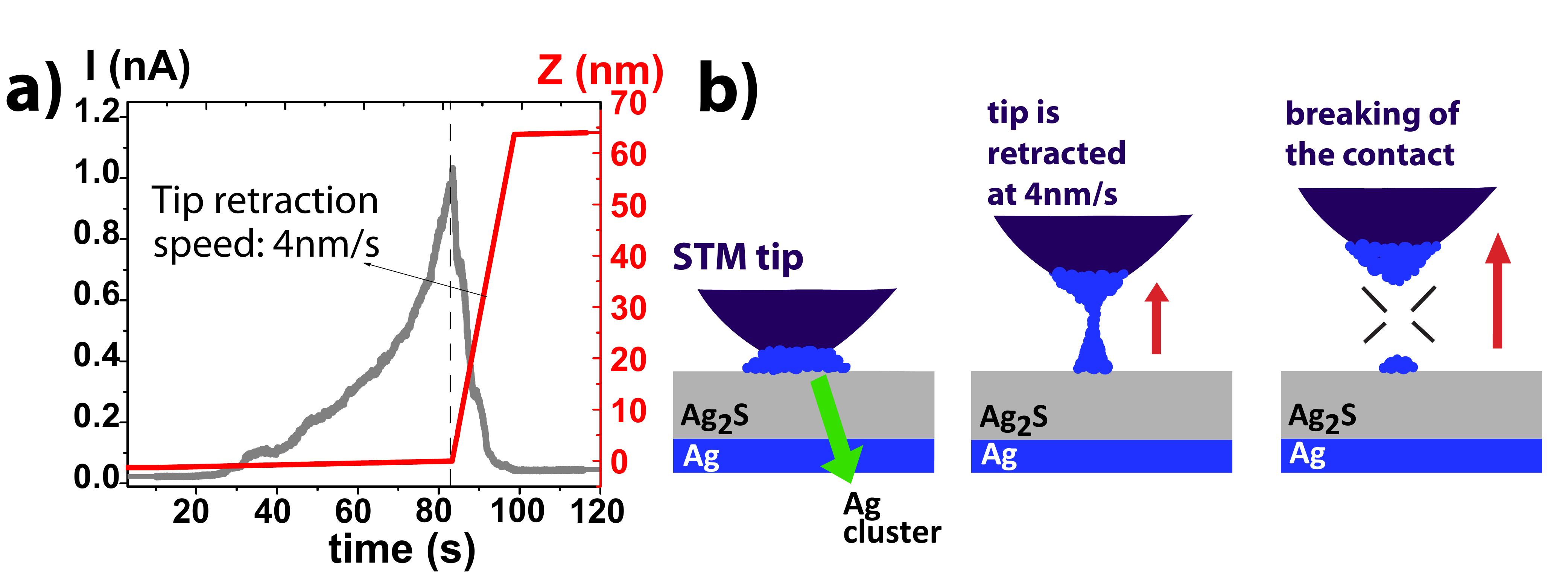}           %data from 061210
      \caption{a. Ag cluster formation induced at +20mV. When the tip current reaches 1 nA, the tip is retracted at a speed of 4nm/s. As a consequence, the tip loses contact with the cluster and the current drops. b. Schematic diagram of the cluster growth and neck formation when the tip is retracted.}
    \label{retract tip}
  \end{center}
\end{figure}

In order to confirm that the tip is indeed in contact with the cluster and that the cluster is metallic, we perform the measurement illustrated in Figure \ref{retract tip}a. \\

For the measurement presented in Figure \ref{retract tip}a, we formed a cluster by applying +20mV sample bias. The current was allowed to increase to 1 nA, and then the tip was retracted at a speed of 4nm/s. It takes approximately 16s for the current to completely drop to the initial value. We observed that the drop of the current starts immediately upon tip retraction from the surface (See line at t $\approx$ 82s in Fig.\ref{retract tip}a). When the cluster is metallic, the drop of the current should be related with a decrease of the cross section area at the Ag$_2$S/Ag-cluster interface. This occurs if the tip is dragging up a large part of the cluster when it is moving away from the surface, as illustrated by the cartoon in Figure \ref{retract tip}b. Figure \ref{retract tip}a shows that in 16 seconds the tip is retracted as much as 64 nm from the sample surface in the Z direction. At much faster retraction rates, the time to break the contact with the cluster is considerably shorter. This experiment clearly shows that the cluster is in contact with the tip and that we can mechanically break this contact by retracting the tip over tens of nanometers.\\

\section{Topography scans}

\begin{figure}[ht]
  \begin{center}
        \includegraphics[width=7cm]{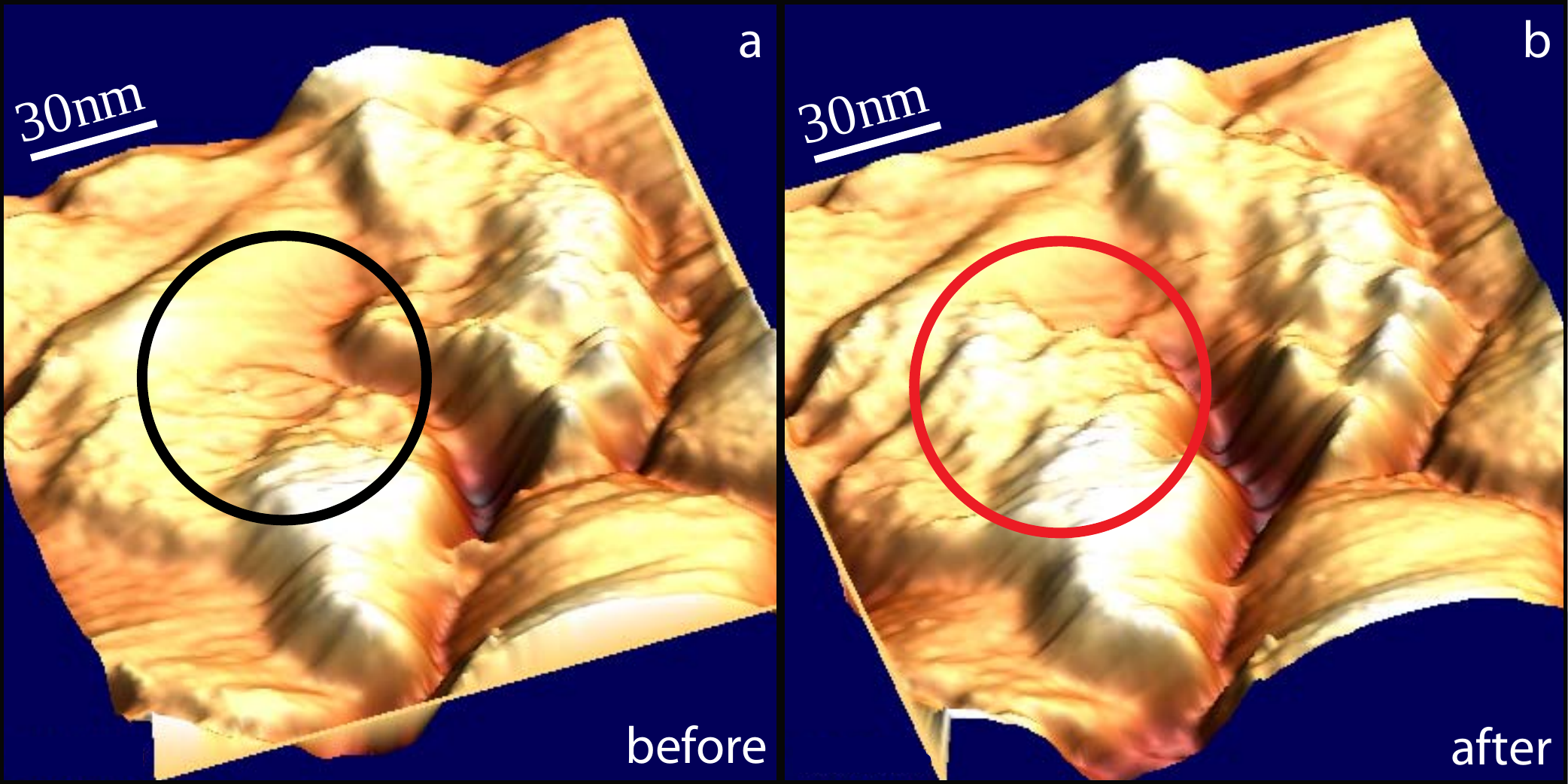}      % data from 180311
        \includegraphics[width=5cm]{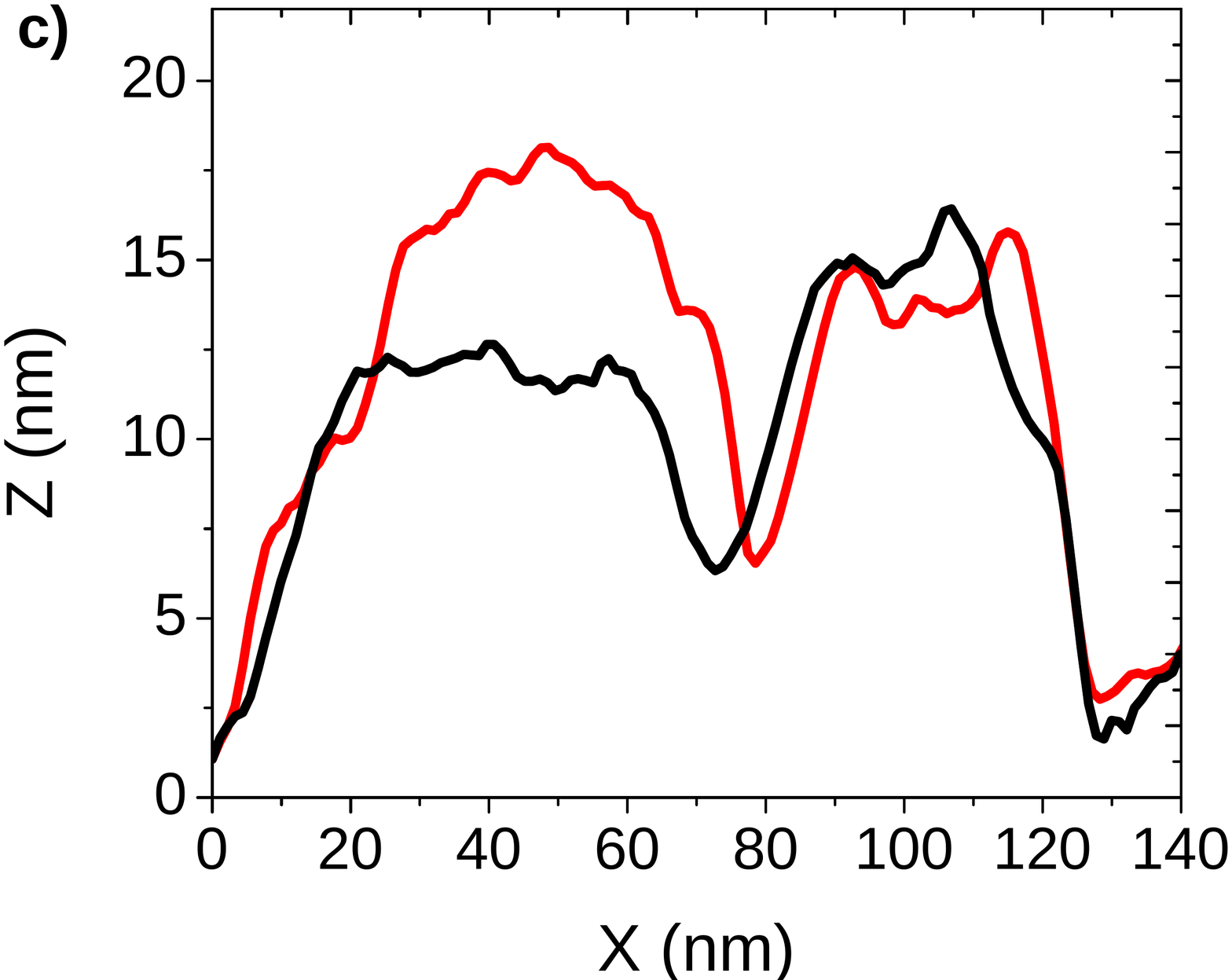}
      \caption{Topography scan a. before and b. after the growth of a cluster. c. Line scan through the remnant of the grown cluster} 
    \label{topo}
  \end{center}
\end{figure}

We scanned the surface of the Ag$_2$S film before and after the growth of a cluster (Figure \ref{topo}). For the scanning we used the tunneling parameters -600mV and 1 nA. The sample was cooled down to T $\approx$ 240 K in order to reduce ion diffusion and prevent complete dissolution of the cluster before the scanning was performed. \\

We scanned an area of 250 x 250 nm (in Figure \ref{topo} we show only an area of 142 x 142 nm) and in the topography image we selected a flat spot on the sample to grow the cluster. The tip was placed above the selected position, and with the feedback on we applied a sample bias of +300mV. This large bias voltage will induce the growth of a large cluster, observed by the large displacement of the Z piezo. This positive bias voltage was applied for 100s. After the 100 s, the tunneling bias was set back (-600mV) and the tip restarted scanning the same area as before cluster growth. \\

Due to the large roughness of the surface of approximately 30nm a slow scan speed needed to be used. The time taken from the moment the cluster was grown to the moment when the area of the cluster was scanned was more than 100s. At room temperature and with -600 mV sample bias the cluster is expected to be fully dissolved back in the Ag$_2$S, as confirmed by our experiments. At 240 K, we observed that after 100s, a small part of the cluster is still visible when the surface is imaged. A line scan through the remnants of the cluster (Figure \ref{topo}c), shows a height difference of 5nm, which is significally smaller than the expected cluster height just after the switch (the measured displacement of the Z piezo while performing the measurement was approximately 100nm). Nevertheless, the two distinct surface morphologies before and after the growth are clearly observed.\\

\section{Discussion}

To understand the nucleation and the evolution of the cluster with time, we performed the following analysis. When a voltage $V$ is applied to the Ag$_2$S sample, between the Ag bottom contact and the Pt tip, the voltage drop over the sample $V_s$ imposes an electrochemical potential difference for the electrons at the two sample boundaries,

\begin{equation}
		-e \: V_s = \tilde{\mu}^{''}_{e} - \tilde{\mu}^{'}_{e}   
\label{eV}   
\end{equation}
with $-e$ the electron charge. Note that the total voltage applied drops partially over the tunneling gap and partially over the sample. Due to the large resistance of the sample, $V$ $\approx$ $V_s$.  \\
 
The electrochemical potential of Ag$^+$-ions and electrons are related to the chemical potential of atoms by,

\begin{equation}
		\mu_{\text{Ag}} = \widetilde{\mu}_{\text{Ag}^{+}} + \widetilde{\mu}_{e}.  
\label{chem-pot}   
\end{equation}  
Then, Eq.(\ref{eV}) can also be written as:

\begin{equation}
		-e \: V_s = \left( \mu^{''}_{\text{Ag}} - \mu^{'}_{\text{Ag}} \right) - \left( \tilde{\mu}^{''}_{\text{Ag}^{+}} - \tilde{\mu}^{'}_{\text{Ag}^{+}} \right)  
\label{eV1}   
\end{equation}

Initially, the Ag$^+$-ions and electrons are distributed uniformly in the sample, thus ${\nabla}{\mu}_{Ag}$ = 0. When the voltage is switched on, the gradients of the electrochemical potentials of electrons and Ag$^+$-ions, will be identical in magnitude but opposite in sign. Opposing currents of ions ($j_{Ag^+}$) and electrons ($j_e$) are therefore set up in the sample.\\

When the top contact is an ion blocking contact (e.g. a Pt tip in contact with the Ag$_2$S sample), the silver ion current ($j_{Ag^+}$) builds a concentration gradient at the Ag$_2$S/Pt-tip interface, until an equilibrium with the electrical potential gradient ${\nabla} \tilde{\mu}_{e}$ is reached, resulting in a steady state and a vanishing current of ions. Increasing the voltage further, increases the Ag concentration gradient, until the supersaturation reaches a critical value and metallic Ag is nucleated inside the Ag$_2$S. This will occur at V$_s$ $>$ 70mV as demonstrated previously \cite{Rickert871983, Morales-Masis2010}.\\

Contrary to a blocking contact, in the present experiment there is initially a tunneling gap between the tip and the sample surface. Therefore ions are not fully blocked at the Ag$_2$S/vacuum interface, and surface nucleation is allowed. Anticipating that the surface nucleation energy is lower than the bulk nucleation energy, a Ag nucleus can be formed at the surface of the Ag$_2$S film at lower Ag saturation. The nucleus is produced by Ag$^+$-ions that are reduced at the surface of the Ag$_2$S film. Initially the nucleus contacts the STM tip without causing a noticeable change in the current due to the large resistance of the sample.\\ 

The further evolution of the cluster radius with time, $r(t)$, is modeled as follows. We define the flux of Ag$^+$-ions, which cross the Ag$_{2}$S/Ag-cluster interface, that are subsequently reduced to metal atoms and added to the nucleus, as $J_{\Omega_t}$. The flux $J_{\Omega_t}$ into the nucleus, therefore, corresponds to a volume change ($\Omega_t$) of the nucleus according to,

\begin{equation}
J_{\Omega_t} \; A(r) \; \Omega_a = \frac{d\Omega_t(r)}{dt}
\label{jAg1}	
\end{equation}   
where $J_{\Omega_t}$ is in units of atoms/m$^2$s, A(r) is the cross section area of the Ag$_{2}$S/Ag-cluster interface, and $\Omega_a$ is the volume per atom. The parameter r is the radius of the cluster, which we assume for simplicity to be cylindrically symmetric.\\

\begin{figure}[ht]
  \begin{center}
  			   \includegraphics[width=7.5cm]{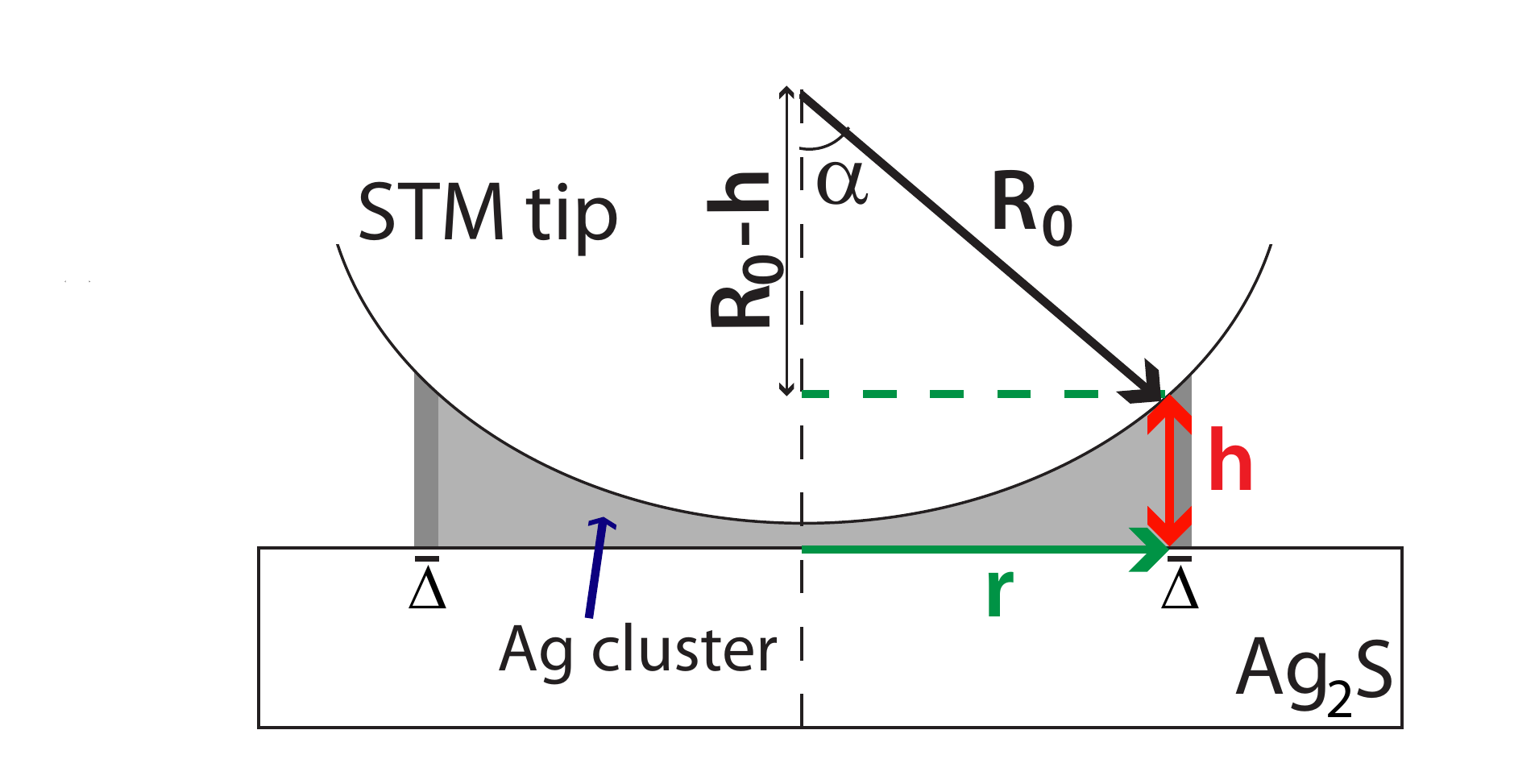}	   
       \caption{Schematic diagram of the tip and sample geometry, used to model the growth of the cluster with time. \textbf{R$_0$} is the tip radius, \textbf{r} is the cluster radius and \textbf{h} is the cluster height. $\Delta$ is the width of the edge ring where the Ag atoms are incorporated into the nucleus.}
    \label{geo}
  \end{center}
\end{figure}

We first solve the equations above for the case of a simplified model of tip and surface geometry. We assume an STM tip with an apex radius R$_0$ and smooth sample surface. We furthermore assume that the cluster grows such that it fills the space between the STM tip and the sample surface. A schematic diagram of this model system is shown in Figure \ref{geo}. \\

Using the parameters defined in Figure \ref{geo}, and assuming that $R_0 >> r$, the height of the cluster can be expressed as $h \approx \frac{r^2}{2 \: R_0}$ and the volume of the cluster as:

\begin{equation}
\Omega_{t}(r) = \frac{\pi \: r^4}{4 \: R_0}
\label{vol}	
\end{equation}

To solve Eq.\ref{jAg1}, let us first analyze $J_{\Omega_t}$. From the measurements presented in Figure \ref{trace50mV150211}, we estimate that the flux of atoms into the cluster is much less than 1$\%$ of the full ion current density in Ag$_2$S (here, the full ion current is obtained from the electronic current using $j_{Ag^+} = 0.1 j_{e}$ \cite{Bonnecaze}). Therefore, when the cluster is growing, the deviation from steady state is very small, and at fixed bias voltage, after allowing for some settling, we can assume that $J_{\Omega_t}$ is constant over time. \\

Because the flux of atoms into the cluster is much smaller than the full ionic current density, we know that the atoms are restricted at the Ag$_{2}$S/Ag-cluster interface. This restriction is the cluster itself. The addition of extra atoms at the cross section area just under the cluster is not likely, since the new atoms would have to work against the atoms in the cluster, to displace then and accumulate in the cluster. At the edge there is no restriction. Thus, we propose that the addition of Ag to the cluster occurs only at the edge of the cluster, this edge having width $\Delta$. This forms an effective surface defined by the edge of the cluster (2 $\pi$ $r(t)$) and a width $\Delta$ of atomic dimensions. The parameter $A(r)$ in Eq.\ref{jAg1} is then given by $A(r) = 2 \: \pi \: r(t) \: \Delta$ with $r(t)$ the radius of the cluster and $\Delta$ constant. \\

We fill in $\Omega_t(r,t)$ and $A(r,t)$ into Eq.\ref{jAg1}, and now we only need to integrate over time. Eq.\ref{jAg1} then results in,\\ 

\begin{equation}
J_{\Omega_t} \; \Omega_a \; t = \frac{r(t)^3}{6 R_0 \Delta}
\label{jAg3}	
\end{equation} 
or
\begin{equation}
r(t) = \left(6 R_0 \Delta \: J_{\Omega_t} \: \Omega_a \: t\right)^{1/3}
\label{r(t)}	
\end{equation}
an expression which describes the evolution of the radius $r$ of the cluster with time. \\

The flux of atoms into the cluster $J_{\Omega_t}$ is proportional to the difference of the chemical potential gradients of Ag atoms and ions in the Ag$_2$S, ${J}_{\Omega_t} \propto \: {\nabla}{\mu}_{Ag} \!-\! {\nabla} \tilde{\mu}_{Ag^+}$ which is expected to be small. To lowest order this flux will be linear in the bias voltage $V_s$. Then Eq. \ref{r(t)} becomes

\begin{equation}
r(V_s,t) \propto \left(R_0 \Delta \: \Omega_a \: V_s \: t\right)^{1/3}
\label{r(t)2}	
\end{equation}

As we will see below, this relation describes the data in some cases very well, but not in general. We note that the power n = 1/3 in Eq.\ref{r(t)2} comes from the specific geometry for the STM tip and cluster presented in Figure \ref{geo}. In fact, Eq.\ref{r(t)2} can be written as a general function $\emph{f}$ of the time t and bias voltage $V_s$, where $\emph{f}$ generalizes the power n of $r(V_s,t)$ defined by the specific shape of the tip and cluster. Then, in general terms,

\begin{equation}
r(V_s, t) = \emph{f} \left( \Psi \: V_s \: t \right)
\label{r(t)gral}	
\end{equation} 
with $ \Psi = R_0 \Delta \: \Omega_a$.\\

Using the expression above and the dependence of the electronic current with the top contact size, we can write a scaling relation for the traces of current I vs time t measured at different bias voltages. \\

As mentioned above, the nucleus formation and growth occur at near-equilibrium conditions, i.e., near steady state, in which case the electronic current is expressed as, \cite{Morales-Masis2010}

\begin{equation}
I(V_s) = K \sigma_{0} \: \frac{k_{B}T}{e}  \left(e^{(eV_s/k_{B}T)} - 1\right) 
\label{I(V)}	
\end{equation}
where $k_{B}$ is Boltzmann's constant, the temperature $T$ = 295K and $K$ = $ \alpha \; r$ describes the contact geometry, where $\alpha$  is a constant of order unity and depends on the size and shape of the top contact. In our case, the top contact is the Ag cluster that is growing at the surface, therefore, $r = r(V_s,t)$.\\

From Eq.\ref{r(t)gral} and Eq.\ref{I(V)}, we obtain the final form,

\begin{equation}
I(V_s,t)  =  \Gamma \: \emph{f}\left(\Psi \: V_s \: t \right) \left(e^{(eV_{s}/k_{B}T)} - 1\right) 
\label{I(V)3}	
\end{equation}
with $\Gamma = \alpha \: \sigma_{0} \: \frac{k_{B}T}{e}$. \\

\begin{figure}[ht]
  \begin{center}
  				 \includegraphics[width=8cm]{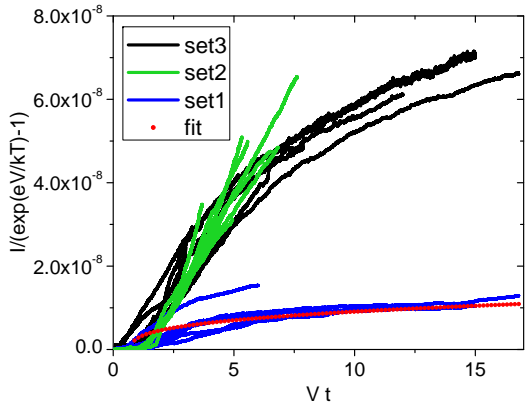}
        \caption{Scaling of three sets of data, measured at different positions on the sample and with different STM tips. The red dotted line is a fit to one of the traces from set 1 with a power law of $V_s t$ of 1/3, as the result obtained with our simplified model (Eq.\ref{r(t)2})}
    \label{scaling}
  \end{center}
\end{figure}

Equation \ref{I(V)3} predicts that if we plot $I(V_s,t)/ \left(e^{(eV_{s}/k_{B}T)} - 1\right)$ versus $V_s \: t $, for one set of data of current versus time traces, for example the set presented in Figure \ref{trace50mV150211}b, the data should collapse onto a single curve. This is confirmed in Figure \ref{scaling} for three different sets of data. Each set of data contains traces of current versus time,  measured at voltages from +20mV to +60mV. Each set of data was measured with a different tip and at a different position on the sample, resulting in a different function $\emph{f}$ and/or proportionally constant $\Gamma$.  For example, the data set 1 was measured with an etched STM tip with a well defined shape, and the data of sets 2 and 3 were measured with a manually cut tip of which the geometry is not well defined. This can explain variations in the power law and the fact that the best scaling with a power law given by Eq.\ref{r(t)2} is obtained for the etched tip (set 1). 

\section{Conclusion}

We have observed the occurrence of two physical processes in Ag$_2$S under the influence of the applied bias voltage: nucleation of a Ag cluster on the surface and, at higher bias, nucleation of a wire inside the Ag$_2$S film. We have found that surface nucleation is possible at very low bias voltages, below 20mV. Nucleation of a nanowire inside the film occurs at bias voltages higher than 70mV at room temperature. At this point it is not possible to decide whether the nucleation is governed strictly by the bias voltage alone, or by the electric field that it causes, as was recently demonstrated for amorphous-crystalline phase change materials \cite{Karpov2008, Karpov2008a, Krebs}. \\

Ag$_2$S forms an interesting model system for the understanding of memristive nano-ionic devices. In nano-ionic devices, the ion transport properties plays a significant role in the determination of retention times, cycling endurance and writing and reading voltages. The type of measurements presented here are important to screen and select future candidate materials and processes for memory resistive devices.  

\section*{Acknowledgements}

We thank H.D. Wiemh$\ddot{\text{o}}$fer for his interest in this project and for discussions. We thank Ruud Tromp for making the UHV-STM system available for this work, Federica Galli for technical support on the STM, Jelmer Wagenaar for discussions. M. Morales-Masis is grateful to Alpana Nayak for support and discussions in the extended visit to Tsukuba. This work is part of the research program of the Dutch
Foundation for Fundamental Research on Matter (FOM) that is financially supported by NWO.

\end{document}